\documentclass[11pt,twoside]{article}
\usepackage{asp2010}

\resetcounters

\def\<<{{\ll}}
\def\>>{{\gg}}

\def\spose#1{\hbox to 0pt{#1\hss}}
\def\ltwig{\mathrel{\spose{\lower 3pt\hbox{$\mathchar"218$}}
     \raise 2.0pt\hbox{$\mathchar"13C$}}}
\def\gtwig{\mathrel{\spose{\lower 3pt\hbox{$\mathchar"218$}}
     \raise 2.0pt\hbox{$\mathchar"13E$}}}
\def\Rstar{R_{\ast}}

\def\Mdot{\dot M}

\def\vinf{V_\infty}

\newcommand{\beq}{\begin{equation}}
\newcommand{\eeq}{\end{equation}}
\newcommand{\beqa}{\begin{eqnarray}}
\newcommand{\eeqa}{\end{eqnarray}}
\def\F21{{_{2}F_{1}}}

\begin{document}

\title{
The Physical Basis of the $L_{x} \sim L_{bol}$ Empirical Law for O-star X-rays
}
\author{Stan Owocki$^1$, Jon Sundqvist$^1$, David Cohen$^2$ and Ken
Gayley$^3$
\affil{$^1$Bartol Research Insitute, 
University of Delaware, Newark,DE 19716 USA}
\affil{$^2$Department of Physics, Swarthmore College, Swarthmore, PA 19081 USA}
\affil{$^3$Department of Physics, University of Iowa, Iowa City, IA
52242 USA}}

\begin{abstract}

X-ray satellites since {\em Einstein} have empirically established that the
X-ray luminosity from single O-stars scales linearly with bolometric
luminosity, $L_{x} \sim 10^{-7} L_{bol}$.  
But straightforward forms of the most
favored model, in which X-rays arise from instability-generated shocks
embedded in the stellar wind, predict a steeper scaling, either with
mass loss rate $L_{x} \sim {\dot M} \sim L_{bol}^{1.7}$
if the shocks are radiative, or
with $L_{x} \sim {\dot M}^{2} \sim  L_{bol}^{3.4}$ if they are adiabatic.
We present here a
generalized formalism that bridges these radiative vs.\ adiabatic
limits in terms of the ratio of the shock cooling length to the local radius.
Noting that the thin-shell instability of radiative shocks should lead
to extensive mixing of hot and cool material, we then propose that the
associated softening and weakening of the X-ray emission can be parameterized
by the cooling length ratio raised to a power $m$, the ``mixing
exponent."  
For physically reasonable values $m \approx 0.4$, this leads to 
an X-ray luminosity $L_{x} \sim {\dot M}^{0.6} \sim L_{bol}$
that matches the empirical scaling. 
We conclude by noting that
such thin-shell mixing may also be important
for X-rays from colliding wind binaries, 
and that
future
numerical simulation studies will be needed to test this thin-shell mixing
{\em ansatz} for X-ray emission.
\end{abstract}

\section{Introduction}

Since the 1970's X-ray satellite missions like {\em Einstein}, {\em Rosat}, and most recently {\em Chandra} and {\em XMM-Newton} have found hot, luminous, O-type stars to be sources of soft ($\ltwig 1$~keV) X-rays, with a  roughly {\em linear} scaling between the X-ray luminosity and the stellar bolometric luminosity, $L_{x} \sim 10^{-7} L_{bol}$ \citep{Gudel09}.
In some systems with harder  (a few keV) spectra and/or higher $L_x$, the observed X-rays have been associated with shock emission in colliding wind binary (CWB) systems, or with magnetically confined wind shocks (MCWS) (see reviews by Corcoran and Gagn\'{e}).
But in putatively single, non-magnetic O-stars, the most favored model is that the X-rays are emitted from Embedded Wind Shocks (EWS) that form from the strong, intrinsic instability (the ``Line-Deshadowing Instability" or LDI)  associated with  the driving of these winds by line-scattering of the star's radiative flux (see review by Sundqvist).

\begin{figure*}
\begin{center}
\includegraphics[scale=.5]{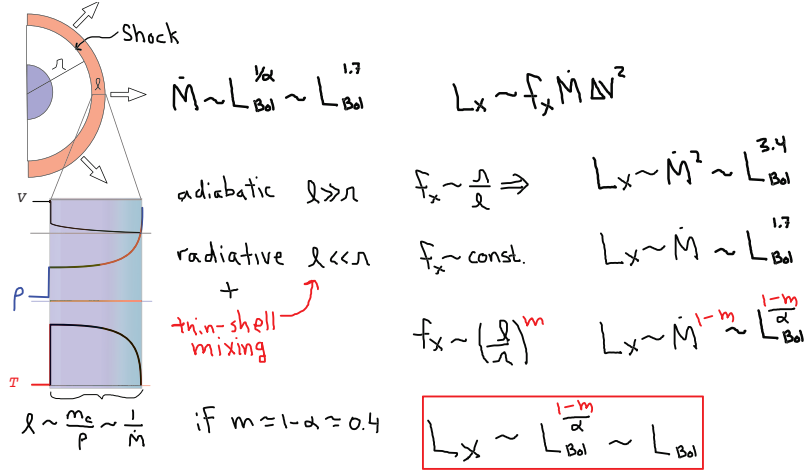}
\caption{Illustration of cooling zone within a wind shock, showing
associated scalings for X-ray luminosity $L_{x}$ with mass loss rate
${\dot M}$ and bolometric luminosity $L_{bol}$, for adiabatic shocks 
with cooling length much larger than the local radius, $\ell \gg r$, 
or radiative shocks with $\ell \ll r$.
Thin-shell mixing of such radiative shocks is posited to lead to a
reduction of the X-ray emitting fraction that scales as a
power-law of cooling length, $f_{x} \sim \ell^{m}$.
For CAK wind index $\alpha$, a mixing exponent $m = 1-\alpha$ leads
to the observationally inferred linear scaling of X-rays with
bolometric luminosity, $L_{x} \sim L_{bol}$.
}
\label{lxlbol-tsm}
\end{center}
\end{figure*}

This LDI can be simply viewed as causing some small ($\ltwig 10^{-3}$) fraction of the wind material to pass through an X-ray emitting EWS, suggesting then that the X-ray luminosity should scale with the wind mass loss rate, $L_x \sim \Mdot$.
But within the standard 
\citet[hereafter CAK]{Castor75}
model for such winds, this mass loss rate increases with luminosity\footnote{For simplicity,  we ignore here secondary scalings, e.g. of luminosity with mass, or wind speed with mass  and radius.} as $\Mdot \sim L_{bol}^{1/\alpha} \sim L_{bol}^{1.7}$, where the latter scaling uses a typical CAK power index $\alpha \approx 0.6$.
This then implies a {\em super-linear} scaling for X-ray to bolometric luminosity, $L_x \sim L_{bol}^{1.7}$, that is too steep to match the observed, near-linear law.

In fact, the above scaling effectively assumes the shocks are {\em radiative}, with a cooling length  that is much smaller than the local radius, $ \ell \ll r$.
In the opposite limit  $\ell \gg r$, applicable to lower-density winds for which shocks  cool by {\em adiabatic} expansion, the shock emission scales with the X-ray {\em emission measure}, 
$EM  \sim \int \rho^2 dV$, leading then to an even steeper scaling of X-ray vs. bolometric luminosity, $L_x \sim \Mdot ^2 \sim L_{bol}^{3.4}$.

Both these scalings ignore  the effect of bound-free absorption of X-rays by the cool, unshocked material that represents the bulk of the stellar wind.
\citet[OC99]{Owocki99b}
 showed that, in an EWS model in which the X-ray emission fraction drops with 
radius, accounting for wind absorption can lead to an observed X-ray luminosity that scales linearly  with $L_{bol}$.
But while modern observations of spectrally resolved X-ray emission profiles by {\em Chandra} and {\em XMM-Newton} do indeed show the expected broadening from EWS, the relatively weak blue-red asymmetry indicates that such absorption effects are modest in even the densest winds
\citep{Cohen10, Cohen11}.
Since many stars following the $L_x$-$L_{bol}$ empirical law have weaker winds that are largely optically thin to X-rays, it now seems clear that absorption cannot explain this broad $L_x$ scaling.

The analysis here examines instead the role of radiative cooling, and associated thin-shell instabilities, in mixing shock-heated material with cooler gas, and thereby reducing and softening the overall X-ray emission. As summarized in figure 1, for a simple parameterization that this mixing reduction scales with a power (the ``mixing exponent'' $m$) of the cooling length, $\ell^m$, we find that the linear  $L_x$-$L_{bol}$ law can be reproduced by assuming $m \approx 0.4$.
To lay the basis for deriving this result in \S 3, 
the next section (\S 2) first introduces a simple bridging form for emission between the radiative
and adiabatic shock limits.

\section{Bridging Law for Adiabatic vs. Radiative Shock Emission}

Building upon the LDI-generated EWS scenario that is  reviewed in this volume by Sundqvist et al.,  
let us model  the associated local X-ray emissivity within the wind as
\beq
\eta_x = C \rho^{2} f_{v}  = C \rho^{2} \frac{f_{q}}{1+r/\ell} \, ,
\label{etaxdef}
\eeq
where $\rho$ is the wind density, $C$ is a constant that depends on
the shock model and atomic physics,
and $f_{v}$ represents a volume filling factor for X-ray emission.
While previous work (e.g., OC99) has often directly parameterized this factor as
following some specified radial function (e.g. power-law), the second equality in eqn.~(\ref{etaxdef})
casts this in terms of a ``bridging law" between the limits for radiative ($\ell \ll r$) and 
adiabatic ($\ell \gg r$) shocks, where $f_q$ now represents some local ``heating
fraction'',  set by LDI-generated EWS.
The  cooling length itself scales as,
\beq
\ell = \frac{m_{c}}{\rho} \equiv \frac{1}{\kappa_{c} \rho} .
\eeq
where the cooling mass column $m_{c}$ depends on the energy of the 
EWS,
and
$\kappa_{c} = 1/m_{c}$ provides a convenient representation with units 
of opacity or mass-absorption coefficient, e.g.\ cm$^{2}$/g.
In the simple model here, we assume that the shock energy is
fixed, and thus that $\kappa_{c}$ is spatially constant.
From eqns.~(18) and (22) of 
\citet{Antokhin04},
we find the
numerical value 
\beq
\kappa_{c} \approx 190 \, E_{kev}^{-2} \, {\rm cm^{2}/g} 
\, ,
\label{kcnum}
\eeq
where $E_{kev}$ is the shock energy in keV.

For X-rays emitted with  photon energy comparable to the shock energy, the bound-free absorption opacity has roughly a similar inverse-square energy dependence, but with a numerical coefficient that is about a factor $\kappa_c/\kappa_{bf} \approx 8$ smaller.
For a wind with mass loss rate $\Mdot$ and flow speed $\vinf$, the  transition from optically thick to thin X-ray emission can be characterized by the unit-optical-depth radius for bound-free absorption,
\beq
R_{1}  \equiv  \frac{\kappa_{bf} \Mdot}{4 \pi \vinf }
\, .
\label{r1def}
\eeq
In direct analogy,
we can define a characteristic {\em adiabatic radius} for transition from radiative to adiabatic cooling of the associated wind shocks,
\beq
R_{a}  \equiv  \frac{\kappa_{c} \Mdot}{4 \pi \vinf } 
\approx
140  R_{\odot} \, \frac{\Mdot_{-6}}{E_{kev}^{2} V_{1000} }
\, ,
\label{radef}
\eeq
where 
$\Mdot_{-6} \equiv \Mdot/10^{-6} M_{\odot}$/yr
and
$V_{1000} \equiv \vinf/1000 \,$ km/s.
For even the densest winds, the X-ray emission onset $R_o \gtwig R_1$, implying, as noted above, that wind absorption is at most a modest effect.
But the stronger coefficient ($\kappa_c/\kappa_{bf} \approx 8$) for radiative cooling means that such winds generally have $R_a  \gg R_o$, implying that most O-star EWS remain radiative well above the wind acceleration region where they are generated
\citep{Zhekov07}.

\section{Thin-Shell Instability and Shock Mixing}

The inherent thinness of radiative shock cooling zones makes them
subject to various thin-shell instabilities 
\citep{Vishniac94}.
These can be expected 
to lead to an unknown, but potentially substantial, level of 
{\em mixing} between cool and hot material.
Since cooler material radiates more efficiently, and in softer
wavebands (toward the UV instead of X-rays), 
such mixing can significantly {\em reduce} the effective X-ray emission.
While there have been some  numerical simulations of the
complex structure that arises from such instabilities 
\citep[e.g.,][]{Walder98},
there unfortunately does not yet appear to be any detailed study of how
this can affect the net X-ray emission.

To characterize the potential mixing effect  on the $L_x$-$L_{bol}$ scaling, let 
us make the plausible {\em ansatz} that the reduction should, for shocks in 
the radiative limit $\ell/r \ll 1$, scale as some
power of the cooling length ratio, $\ell/r$.
To ensure that the mixing effect goes away in the adiabatic limit, we 
can (much as in eqn.~\ref{etaxdef}) assume a  simple `bridging law' scaling for a
``mixing reduction factor'' for X-rays,
\beq
f_{xm} = \frac{1}{(1+r/\ell)^{m}}
\, ,
\eeq where the mixing exponent $m>0$.
To account for mixing within this model, we thus simply multiply the
 X-ray emissivity $\eta_x$  in eqn.~(\ref{etaxdef}) by this mixing factor $f_{xm}$.

As a specific, simple model, let us next also assume that, beyond some onset radius $R_o$, 
the X-ray heating fraction declines as power-law in radius, $f_q (r) = f_{qo} (R_o/r)^q$.
%
Neglecting absorption, the X-ray luminosity can then be obtained from spherical volume 
integration of this X-ray emission, 
\def\cq{C_{q}}
\beq
L_{x} = 4 \pi  C \int_{R_o}^\infty  f_{xm} f_v  \rho^2 \, r^2 dr
=
4 \pi \cq
\left ( \frac{\Mdot}{ 4 \pi \vinf} \right )^{2} 
\int_{R_{o}}^{\infty} 
\frac{dr}{
r^{q} \, 
\left ( r  w + R_{a} \right )^{1+m} \,
(r w)^{1-m}
}
\, ,
\label{lxgen}
\eeq
where
 $\cq \equiv C f_{qo} R_{o}^{q}$,
 and $w(r) \equiv V(r)/\vinf $ is the scaled wind speed.
For the standard $\beta=1$ velocity law,
we have $rw = r - \Rstar$, which even for
 general values of $q$ and $m$ allows 
 analytic integration of (\ref{lxgen}) in terms of the Appell Hypergeometric function.
As a specific example, for shock heating that declines with inverse radius ($q=1$),
direct numerical evaluation shows that the total X-ray luminosity is well approximated 
by a simple bridging  law between the radiative and  adiabatic limits,
\beqa
L_{x} 
&\approx& 
4 \pi 
\cq
 \,  \left ( \frac{\Mdot}{ 4 \pi \vinf  R_\ast}  \right )^{2} 
\left \{ \frac{1}{R_o/R_\ast-1} + \ln{(1-R_\ast/R_o )}  \right \}
\, 
\left [
\frac{R_o}
{R_o+R_{a}/ (1+m)}
\right ]^{1+m}
\, 
\\
&\propto& ~ \left ( \frac{\Mdot}{ \vinf } \right )^{2} ~~	~~~~ ; ~~ R_a \ll R_o
\\
&\propto& ~ \left (\frac{\Mdot}{ \vinf } \right )^{1-m} ~~ ; ~~ R_a \gg R_o \, ,
\label{lxq1sm}
\eeqa
where the curly bracket term follows from straightforward integration of (7) for $R_a \rightarrow 0$, 
in which case the square bracket term just becomes unity;
the latter scalings follow from limit evaluations of this square bracket, using the definition of $R_a$ in eqn.~(\ref{radef}).
The transition $R_a \approx R_o$ marks a kind of ``sweet spot''  for conversion of shock energy into  X-ray emission; for lower density winds ($R_a < R_o$) much of that energy is lost to adiabatic expansion, while for higher density winds ($R_a > R_o$), it is lost to thin-shell mixing.

In very dense winds with optically thick X-ray emission and so $R_1 > R_o$, one can approximately account for the associated wind absorption through an ``exospheric" approach (OC99) in which $R_1$ simply replaces $R_o$ as the lower bound for the integral in (\ref{lxgen}), and thus also in (8).
Since $R_a/R_1 = \kappa_c/\kappa_{bf} \approx 8 \gg 1$, the square bracket term just becomes a fixed constant, independent of $\Mdot$. Moreover, expansion of the curly bracket term now also makes the overall $L_x$ scaling {\em independent} of $\Mdot$ for this $f_q \sim 1/r$ ($q=1$) emission case in the dense wind limit, $R_a > R_1 \gg R_o$.

\section{Summary and Future Work}

The key result of this paper is that, in the common case of moderately dense winds with radiative shocks  ($R_a > R_o$), thin-shell mixing can lead to a {\em sub-linear} scaling 
of the X-ray luminosity with the mass-loss rate, 
$L_{x} \sim (\Mdot/\vinf)^{1-m}$.
For a quite reasonable mixing exponent value $ m \approx 1-\alpha \approx 0.4$, this then gives roughly the {\em linear} $L_x$-$L_{bol}$ law that is empirically observed for O-star X-rays.

However we note that a similar mixing analysis could also be applied to model X-ray emission from colliding wind binaries, and their $L_x$ scaling with orbital separation. Wide binaries with adiabatic shocks should still follow the usual inverse distance scaling, as directly confirmed by observations of multi-year-period elliptical systems like WR140 and $\eta$~Carianae (see review by Corcoran). But in close, short (day to week) period binaries with radiative shocks, mixing could reduce and limit the effective X-ray emission from the wind collision, and thus help explain why such systems often hardly exceed the $L_x \approx 10^{-7} L_{bol}$ scaling found for 
single stars (see review  by Gagn\'{e}).

Finally, in addition to exploring such effects in colliding wind binaries,
an overriding priority for future work should be to carry out detailed simulations of the general effect of thin-shell mixing on X-ray emission, and specifically to examine the validity of this
 mixing-exponent {\em ansatz} for modeling the resulting scalings for X-ray luminosity.

\acknowledgements{This work was carried out with support from NASA ATP grant NNX11AC40G.}

\bibliography{OwockiS}

\begin{thebibliography}{}
\expandafter\ifx\csname natexlab\endcsname\relax\def\natexlab#1{#1}\fi
\expandafter\ifx\csname url\endcsname\relax
  \def\url#1{\texttt{#1}}\fi
\expandafter\ifx\csname urlprefix\endcsname\relax\def\urlprefix{URL }\fi
\providecommand{\eprint}[2][]{\url{#2}}

\bibitem[{{Antokhin} et~al.(2004){Antokhin}, {Owocki}, \& {Brown}}]{Antokhin04}
{Antokhin}, I.~I., {Owocki}, S.~P., \& {Brown}, J.~C. 2004, \apj, 611, 434

\bibitem[{{Castor} et~al.(1975){Castor}, {Abbott}, \& {Klein}}]{Castor75}
{Castor}, J.~I., {Abbott}, D.~C., \& {Klein}, R.~I. 1975, \apj, 195, 157

\bibitem[{{Cohen} et~al.(2011){Cohen}, {Gagn{\'e}}, {Leutenegger}, {MacArthur},
  {Wollman}, {Sundqvist}, {Fullerton}, \& {Owocki}}]{Cohen11}
{Cohen}, D.~H., {Gagn{\'e}}, M., {Leutenegger}, M.~A., {MacArthur}, J.~P.,
  {Wollman}, E.~E., {Sundqvist}, J.~O., {Fullerton}, A.~W., \& {Owocki}, S.~P.
  2011, \mnras, 415, 3354. \eprint{1104.4786}

\bibitem[{{Cohen} et~al.(2010){Cohen}, {Leutenegger}, {Wollman}, {Zsarg{\'o}},
  {Hillier}, {Townsend}, \& {Owocki}}]{Cohen10}
{Cohen}, D.~H., {Leutenegger}, M.~A., {Wollman}, E.~E., {Zsarg{\'o}}, J.,
  {Hillier}, D.~J., {Townsend}, R.~H.~D., \& {Owocki}, S.~P. 2010, \mnras, 405,
  2391. \eprint{1003.0892}

\bibitem[{{G{\"u}del} \& {Naz{\'e}}(2009)}]{Gudel09}
{G{\"u}del}, M., \& {Naz{\'e}}, Y. 2009, \aapr, 17, 309. \eprint{0904.3078}

\bibitem[{{Owocki} \& {Cohen}(1999)}]{Owocki99b}
{Owocki}, S.~P., \& {Cohen}, D.~H. 1999, \apj, 520, 833.
  \eprint{arXiv:astro-ph/9901250}

\bibitem[{{Vishniac}(1994)}]{Vishniac94}
{Vishniac}, E.~T. 1994, \apj, 428, 186. \eprint{arXiv:astro-ph/9306025}

\bibitem[{{Walder} \& {Folini}(1998)}]{Walder98}
{Walder}, R., \& {Folini}, D. 1998, \aap, 330, L21

\bibitem[{{Zhekov} \& {Palla}(2007)}]{Zhekov07}
{Zhekov}, S.~A., \& {Palla}, F. 2007, \mnras, 382, 1124. \eprint{0708.0085}

\end{thebibliography}

\end{document}